\documentclass{svproc}
\usepackage{graphicx}
\usepackage{color, soul}
\usepackage{amsmath, amsfonts, amssymb}
\usepackage{url}

\begin{document}
\mainmatter              % start of a contribution
\title{Minimum entropy stochastic block models neglect edge distribution heterogeneity.}
\titlerunning{Minimum entropy stochastic block models}  % abbreviated title (for running head)
\author{Louis Duvivier\inst{1} \and C\'eline Robardet\inst{1} \and R\'emy Cazabet\inst{2} }
\authorrunning{Louis Duvivier et al.} % abbreviated author list (for running head)
%\tocauthor{Ivar Ekeland, Roger Temam, Jeffrey Dean, David Grove, Craig Chambers, Kim B. Bruce, and Elisa Bertino}
%
\institute{
    Univ Lyon, INSA Lyon, CNRS, LIRIS UMR5205, F-69621 France \\
    \email{louis.duvivier@insa-lyon.fr}, \email{celine.robardet@insa-lyon.fr}, \\
    \and
    Univ Lyon, Universit\'e Lyon 1, CNRS, LIRIS UMR5205, F-69622 France \\
    \email{remy.cazabet@univ-lyon1.fr}
}

\maketitle

\begin{abstract}
The statistical inference of stochastic block models as emerged as a mathematicaly principled method for identifying communities inside networks. Its objective is to find the node partition and the block-to-block adjacency matrix of maximum likelihood \textit{i.e.} the one which has most probably generated the observed network. In practice, in the so-called microcanonical ensemble, it is frequently assumed that when comparing two models which have the same number and sizes of communities, the best one is the one of minimum entropy \textit{i.e.} the one which can generate the less different networks. In this paper, we show that there are situations in which the minimum entropy model does not identify the most significant communities in terms of edge distribution, even though it generates the observed graph with a higher probability.

\end{abstract}

Since the seminal paper by Girvan and Newman \cite{girvan2002community}, a lot of work has been devoted to finding community structure in networks \cite{fortunato2016community}. The objective is to exploit the heterogeneity of connections in graphs to partition its nodes into groups and obtain a coarser description, which may be simpler to analyze. Yet, the absence of a universally accepted formal definition of what a community is has favored the development of diverse methods to partition the nodes of a graph, such as the famous modularity function \cite{newman2004finding}, and the statistical inference of a stochastic block model \cite{hastings2006community}.

This second method relies on the hypothesis that there exists an original partition of the nodes, and that the graph under study was generated by picking edges at random with a probability that depends only on the communities to which its extremities belong. The idea is then to infer the original node partition based on the observed edge distribution in the graph. This method has two main advantages with respect to modularity maximization: first, it is able to detect non-assortative connectivity pattern, \textit{i.e.} groups of nodes that are not necessarily characterized by an internal density higher than the external density, and second it can be performed in a statistically significant way, while it has been shown that modularity may detect communities even in random graphs \cite{guimera2004modularity}. In particular, a bayesian stochastic blockmodeling approach has been developed in \cite{peixoto2017nonparametric}, which finds the most likely original partition for a SBM with respect to a graph by maximizing simultaneously the probability to choose this partition and the probability to generate this graph, given the partition.

To perform the second maximization, this method assumes that all graphs are generated with the same probability and it thus searches a partition of minimal entropy, in the sense that the cardinal of its microcanonical ensemble (\textit{i.e.} the number of graphs the corresponding SBM can theoretically generate \cite{cimini2019statistical}) is minimal, which is equivalent to maximizing its likelihood\cite{peixoto2012entropy}. In this paper, we show that even when the number and the size of the communities are fixed, the node partition which corresponds to the sharper communities is not always the one with the lower entropy. We then demonstrate that when community sizes and edge distribution are heterogeneous enough, a node partition which places small communities where there are the most edges will always have a lower entropy. Finally, we illustrate how this issue implies that such heterogeneous stochastic block models cannot be identified correctly by this model selection method and discuss the relevance of assuming an equal probability for all graphs in this context.

\section{Entropy based stochastic block model selection}
\label{sec_entropy_selection}

The stochastic block model is a generative model for random graphs. It takes as parameters a set of nodes $V = [1;n]$ partitioned in $p$ blocks (or communities) $C = (c_i)_{i \in [1;p]}$ and a block-to-block adjacency matrix $M$ whose entries correspond to the number of edges between two blocks. The corresponding set of generable graphs $G = (V,E)$ with weight matrix $W$ is defined as:

\[
\Omega_{C,M} = \left\{G \mid \forall c_1, c_2 \in C, \sum_{i \in c_i, j \in c_j} W_{(i, j)} = M_{(c_1, c_2)} \right\}
\]

It is called the microcanonical ensemble (a vocabulary borrowed to statistical physics \cite{cimini2019statistical}) and it can be refined to impose that all graphs are simple, undirected (in which case $M$ must be symmetric) and to allow or not self loops. In this paper we will consider multigraphs with self loops, because they allow for simpler computations. Generating a graph with the stochastic block model associated to $C,M$ amounts to drawing at random $G \in \Omega_{C,M}$. The probability distribution $\mathbb{P}[G|C,M]$ on this ensemble is defined as the one which maximizes Shanon's entropy 
\[S = \underset{G \in \Omega_{C,M}}{\sum}\mathbb{P}[G|C,M] \times \mathrm{ln}(\mathbb{P}[G|C,M])
\]
In the absence of other restriction, the maximum entropy distribution is the flat one:
\[
\mathbb{P}[G|C,M] = \frac{1}{|\Omega_{C,M}|}
\]
whose entropy equals $S = \mathrm{ln}(|\Omega_{C,M}|)$. It has been computed for different SBM flavours in \cite{peixoto2012entropy}. It measures the number of different graphs a SBM can generate with a given set of parameters. The lower it is, the higher the probability to generate any specific graph $G$.

On the other hand, given a graph $G = (V,E)$, with a weight matrix $W$, it may have been generated by many different stochastic block models. For any partition $C = (c_i)_{i \in [1;p]}$ of $V$, there exists one and only one matrix $M$ such that $G \in \Omega_{C,M}$, and it is defined as:
\[
    \forall c_1, c_2 \in C, M_{(c_1, c_2)} = \sum_{i \in c_1, j \in c_2} W_{(i,j)}
\]
Therefore, when there is no ambiguity about the graph $G$, we will consider indifferently a partition and the associated SBM in the following.

The objective of stochastic block model inference is to find the partition $C$ that best describes $G$. To do so, bayesian inference relies on the Bayes theorem which stands that: 

\begin{equation}
  \mathbb{P}[C,M|G] = \frac{\mathbb{P}[G|C,M] \times \mathbb{P}[C,M]}{\mathbb{P}[G]}
  \label{bayes}
\end{equation}

As $\mathbb{P}[G]$ is the same whatever $C$, it is sufficient to maximize $\mathbb{P}[G|C,M] \times \mathbb{P}[C,M]$. The naive approach which consists in using a maximum-entropy uniform prior distribution for $\mathbb{P}[C,M]$ simplifies the computation to maximizing directly $\mathbb{P}[G|C]$ (the so called likelihood function) but it will always lead to the trivial partition $\forall i \in V, c_i = \{i\}$, which is of no use because the corresponding SBM reproduces $G$ exactly: $M = W$ and $\mathbb{P}[G|C] = 1$. To overcome this overfitting problem, another prior distribution was proposed in \cite{peixoto2017bayesian}, which assigns lower probabilities to the partitions with many communities. Yet, when comparing two models $C_1, M_1$ and $C_2, M_2$ with equal probability, the one which is chosen is still the one minimizing $|\Omega_{C,M}|$ or equivalently the entropy $S = \mathrm{ln}(|\Omega_{C,M}|)$, as logarithm is a monotonous function.

\section{The issue with heavily populated graph regions}
\label{sec_high_density_zoom}

In this paper, we focus on the consequence of minimizing the entropy to discriminate between node partitions. To do so, we need to work on a domain of partitions on which the prior distribution is uniform. As suggested by \cite{peixoto2017bayesian}, we restrict ourselves to finding the best partition when the number $p$ and the sizes $(s_i)_{i \in [1;p]}$ of communities are fixed because in this case, both $P[C]$ and $P[M|C]$ are constant. This is a problem of node classification, and in this situation the maximization of equation \ref{bayes} boils down to minimizing the entropy of $\Omega_{C,M}$, which can be written as:
\[
S = \sum_{i,j \in [1;p]}\mathrm{ln}\left[{s_is_j + M_{(i,j)} - 1 \choose M_{(i,j)}}\right]
\]
as shown in \cite{peixoto2012entropy}.

\begin{figure}[t]
  \includegraphics[width=0.48\textwidth]{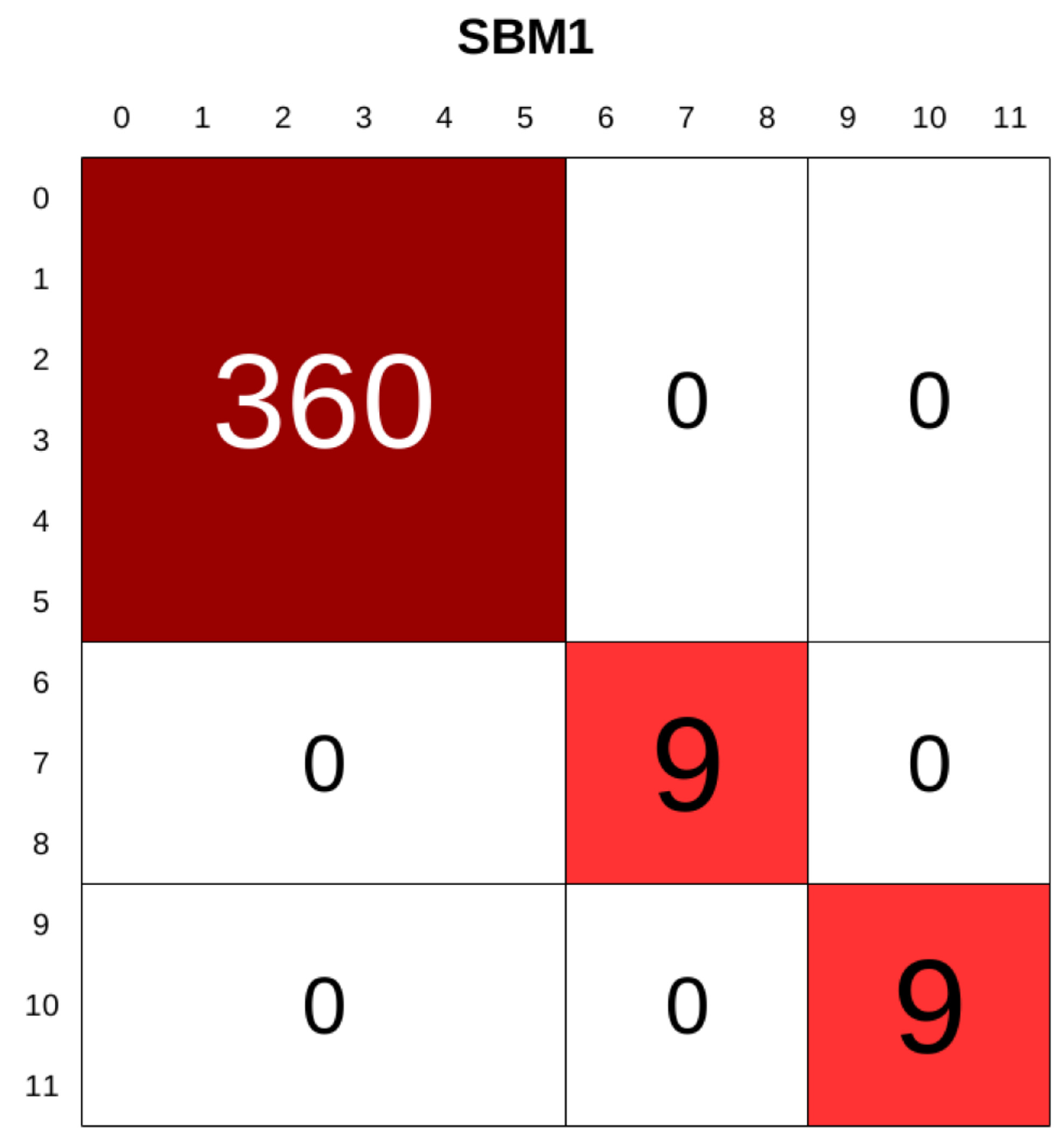}
  \hfill
  \includegraphics[width=0.48\textwidth]{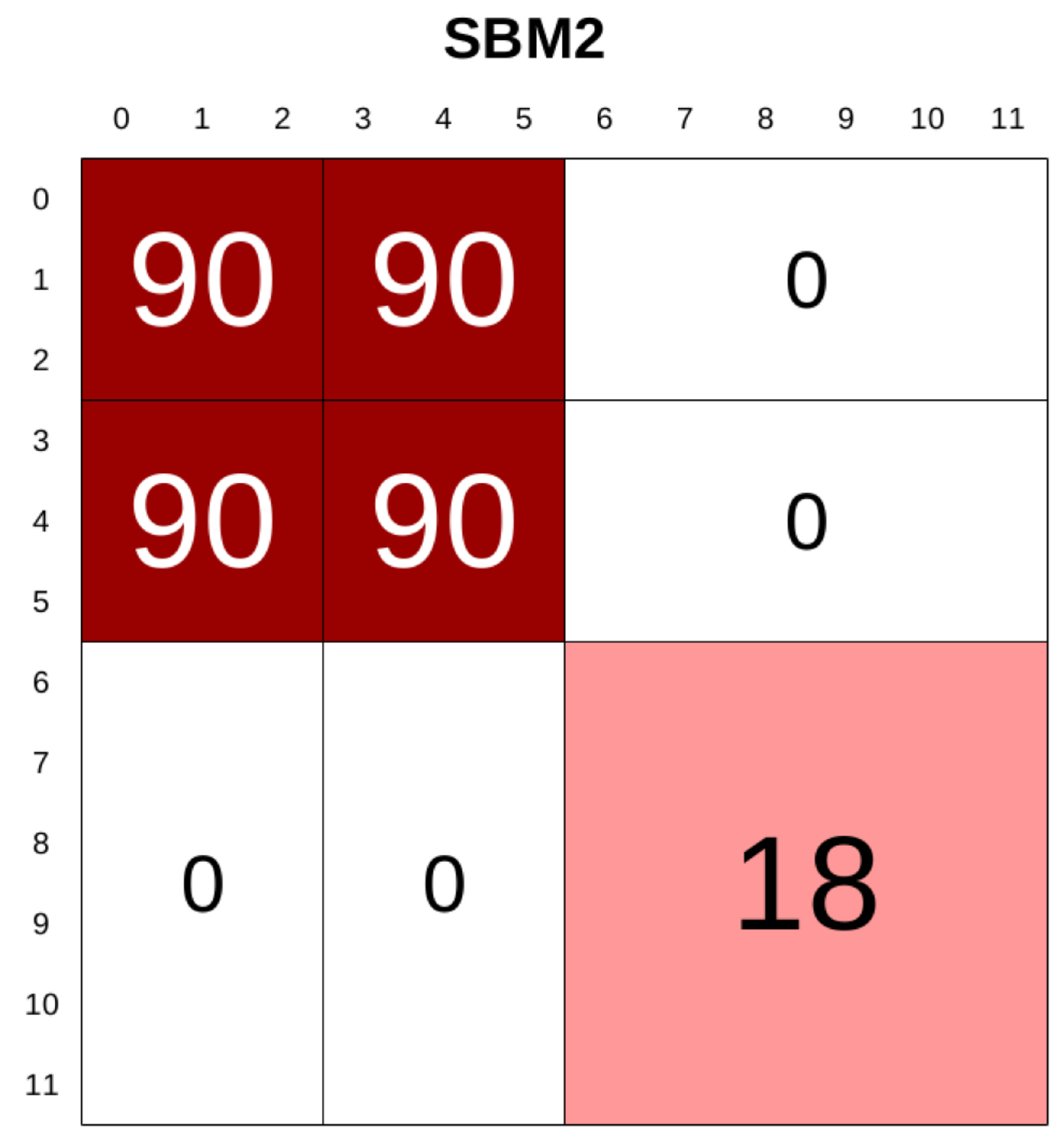}
  \caption{\label{SBM}\textbf{Block-to-block adjacency matrices of two overlapping stochastic block models.} Even though the communities of $SBM_1$ are better defined, $SBM_2$ can generate less different graphs and thus generates them with higher probability.}
\end{figure}

Yet, even within this restricted domain ($p$ and $(s_i)_i$ are fixed), the lower entropy partition for a given graph $G$ is not always the one which corresponds to the sharper communities. To illustrate this phenomena, let's consider the stochastic block models whose matrices $M$ are shown on figure \ref{SBM}, and a multigraph $G \in \Omega_{SBM_1} \cap \Omega_{SBM_2}$. 
\begin{itemize}
\item $SBM_1$ corresponds to $C_1 = \{c_1^a: \{0,1,2,3,4,5\}, c_1^b: \{6,7,8\}, c_1^c: \{9,10,11\}\}$
\item $SBM_2$ corresponds to $C_2 = \{c_2^a: \{0,1,2\}, c_2^b: \{3,4,5\}, c_2^c: \{6,7,8,9,10,11\}\}$. 
\end{itemize}
As $G \in \Omega_{SBM_1} \cap \Omega_{SBM_2}$, it could have been generated using $SBM_1$ or $SBM_2$. Yet, the point of inferring a stochastic block model to understand the structure of a graph is that it is supposed to identify groups of nodes (blocks) such that the edge distribution between any two of them is homogeneous and characterized by a specific density. From this point of view $C_1$ seems a better partition than $C_2$:
\begin{itemize}
    \item The density of edges inside and between $c_2^a$ and $c_2^b$ is the same (10), so there is no justification for dividing $c_1^a$ in two.
    \item On the other hand, $c_1^b$ and $c_1^c$ have an internal density of 1 and there is no edge between them, so it is logical to separate them rather than merge them into $c_2^c$.
\end{itemize}
Yet, if we compute the entropy of $SBM_1$ and $SBM_2$:
\[
  S_1 = \mathrm{ln}\left[{395 \choose 360}\right] + 2 \times \mathrm{ln}\left[{17 \choose 9}\right] = 136
  \]
  
\[
  S_2 = \mathrm{ln}\left[{53 \choose 18}\right] + 4 \times \mathrm{ln}\left[{98 \choose 90}\right] = 135
\]
The entropy of $SBM_2$ is lower and thus partition $C_2$ will be the one selected. Of course, as $|\Omega_{SBM_2}| < |\Omega_{SBM_1}|$, the probability to generate $G$ with $SBM_2$ is higher than the probability to generate it with $SBM_1$. But this increased probability is not due to a better identification of the edge distribution heterogeneity, it is a mechanical effect of imposing smaller communities in the groups of nodes which contain the more edges, even if their distribution is homogeneous. Doing so reduces the number of possible positions for each edge and thus the number of different graphs the model can generate.

\begin{figure}[t]
  \includegraphics[width=0.49\textwidth]{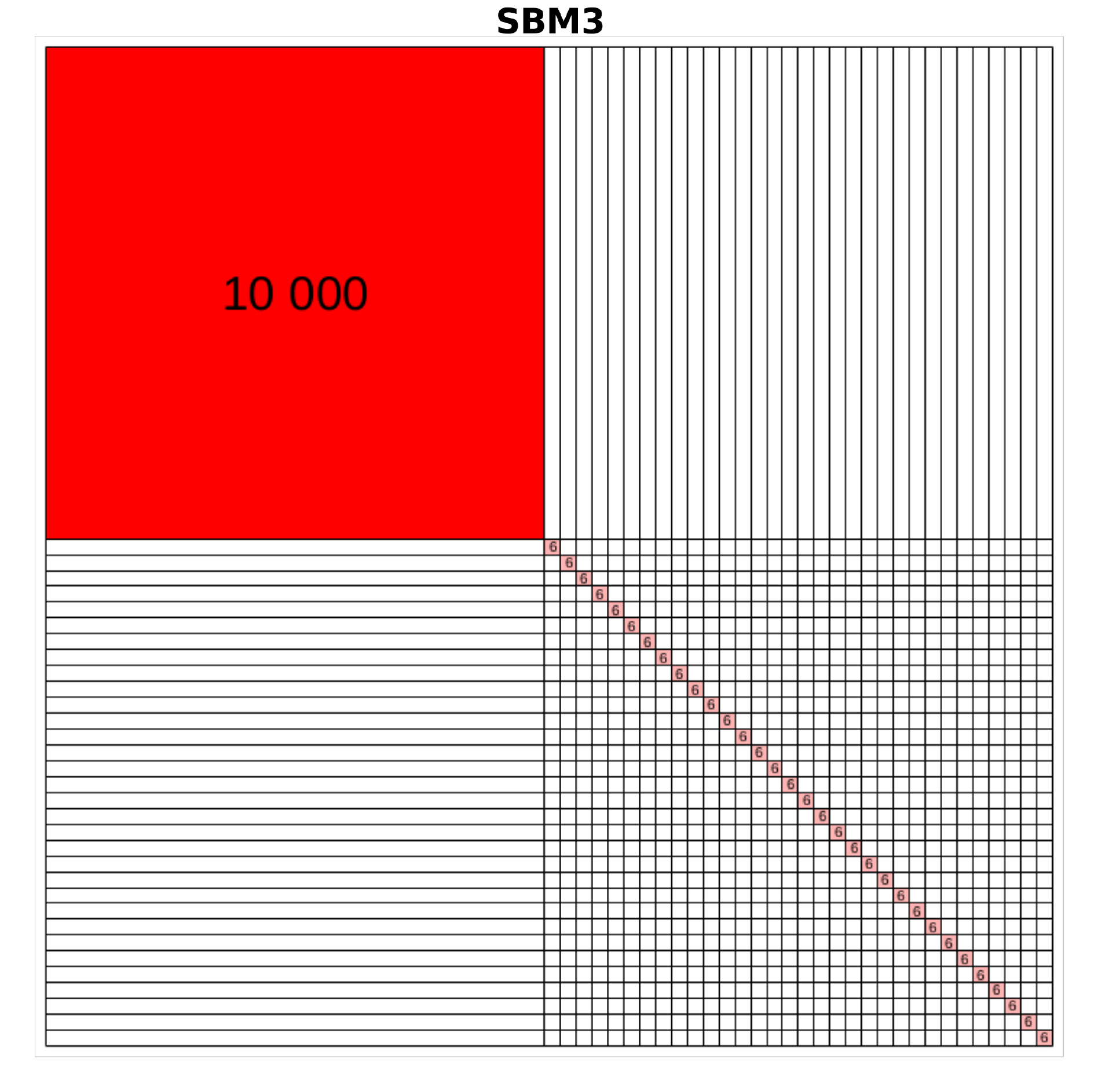}
  \hfill
  \includegraphics[width=0.49\textwidth]{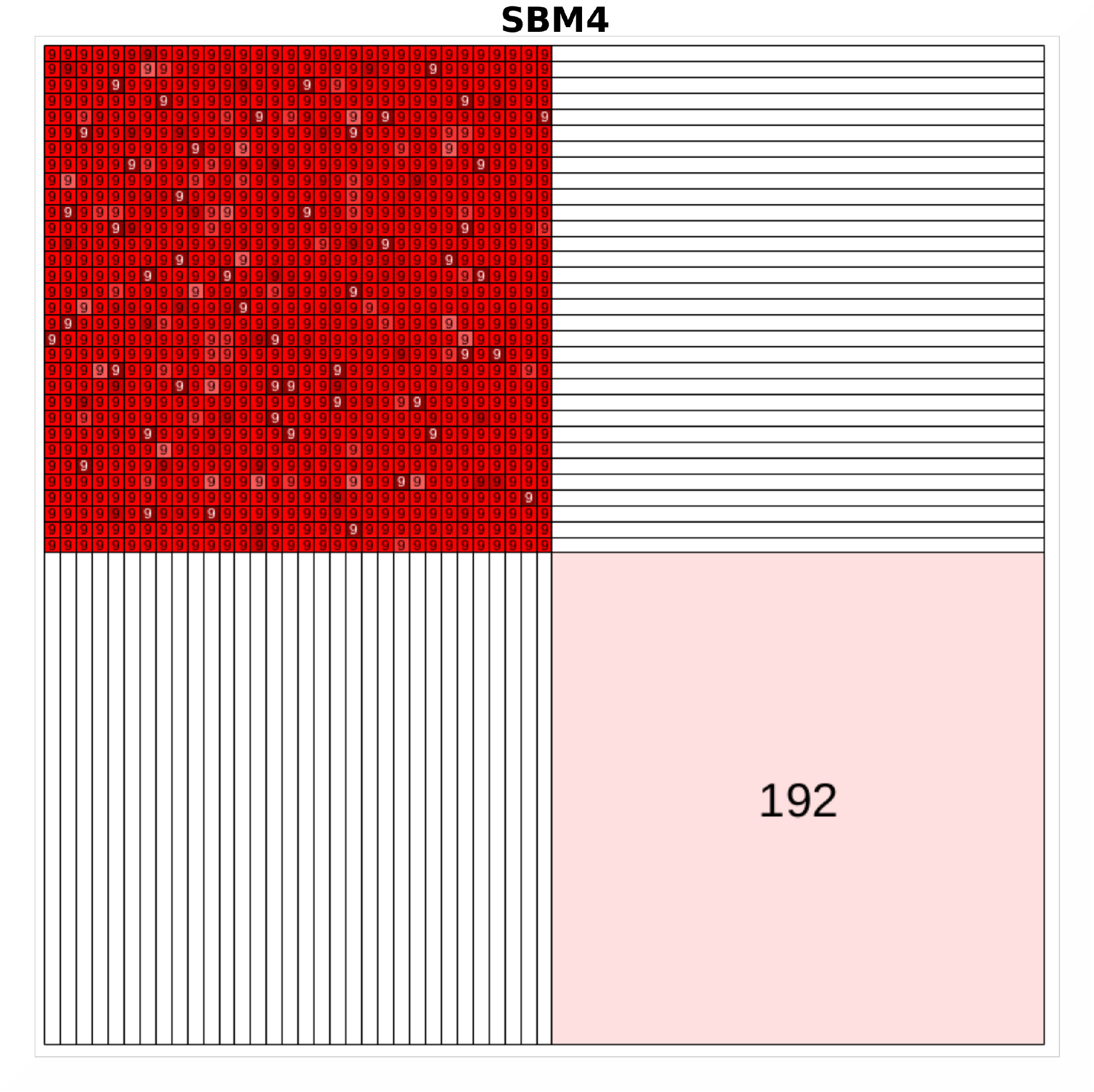}
  \caption{\label{SBM2}\textbf{Block-to-block adjacency matrices of two overlapping stochastic block models with lower densities.} Once again, even though $SBM_3$ has better defined communities, $SBM_4$ is more likely a model for graphs $G \in \Omega_{SBM_3} \cap \Omega_{SBM_4}$}
\end{figure}

This problem can also occur with smaller densities, as illustrated by the stochastic block models whose block-to-block adjacency matrices are shown on figure \ref{SBM2}. $SBM_3$, defined as one community of 128 nodes and density 0.6 and 32 communities of 4 nodes and density 0.4 has an entropy of 17851. $SBM_4$ which merges all small communities into one big and splits the big one into 32 small ones has an entropy of 16403.

\clearpage

\section{The density threshold}
\label{sec_density_threshold}

\begin{figure}[]
  \includegraphics[width=0.48\textwidth]{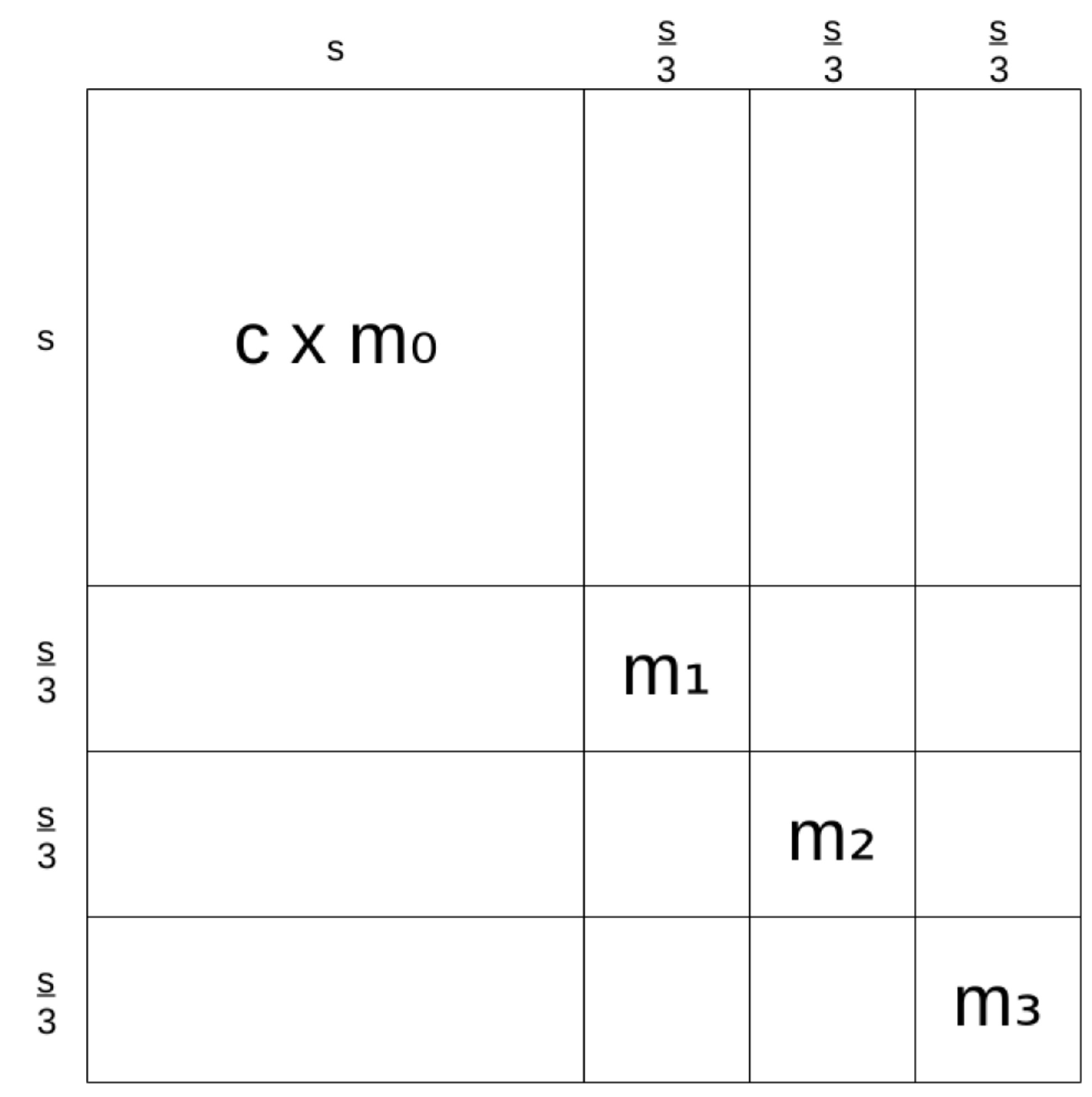}
  \hfill
  \includegraphics[width=0.48\textwidth]{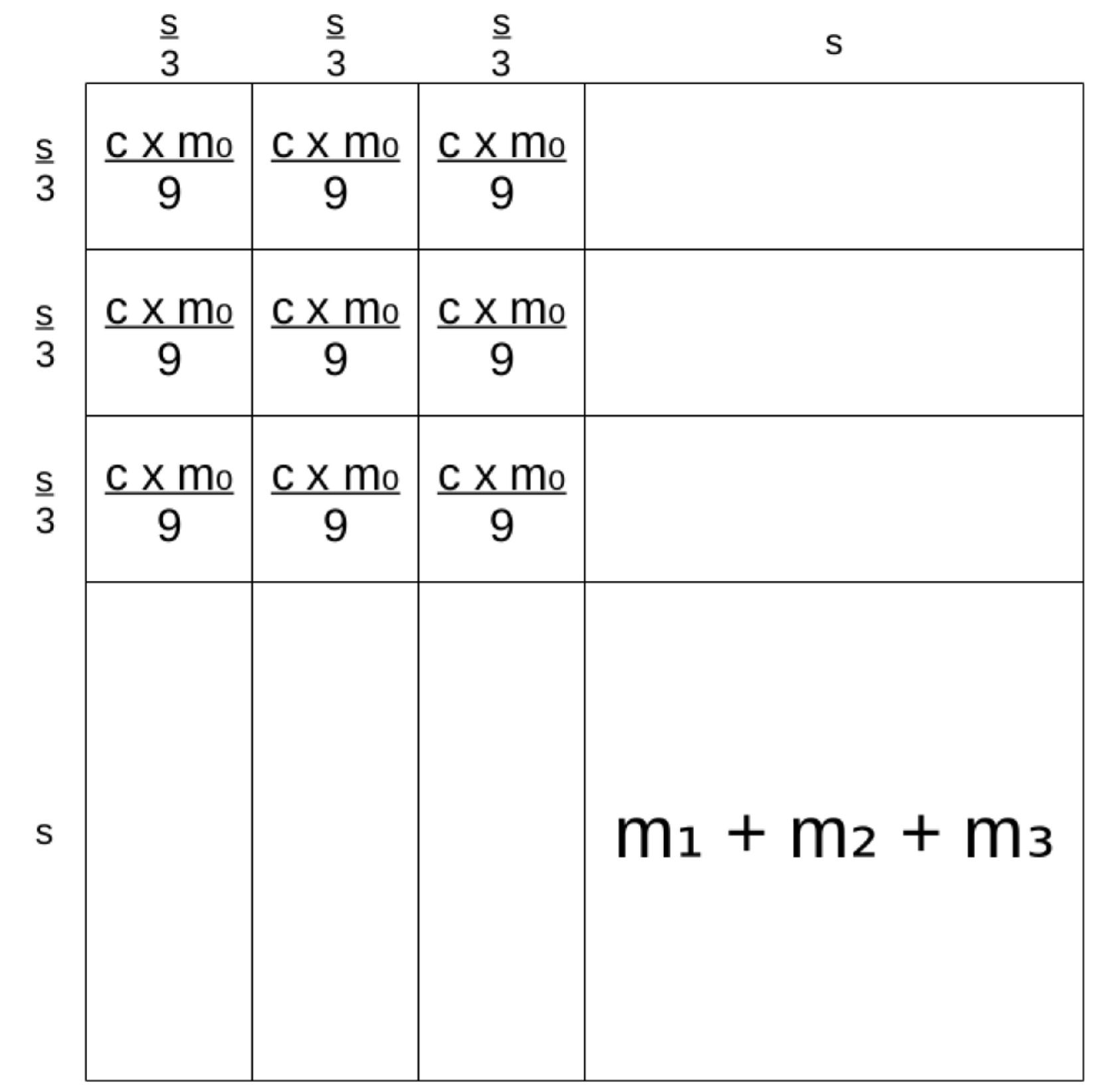}
  \caption{\label{SBM3}\textbf{Theoretical pair of stochastic block models.} The right-side partition splits the big community in $q = 3$ small ones and merges the small communities in one big.}
\end{figure}

More generally, let's consider a SBM $(C_1, M_1)$ with one big community of size $s$, containing $c \times m_0$ edges and $q$ small communities of size $\frac{s}{q}$ containing $(m_i)_{i \in [1;q]}$ edges each, as illustrated on figure \ref{SBM3}. Its entropy is equal to:
\[
S_1(c) = \mathrm{ln}\left[{s^2 + c \times m_0 - 1 \choose c \times m_0}\right] + \sum_{i=1}^q \mathrm{ln}\left[{\frac{s^2}{q^2} + m_i - 1 \choose m_i}\right]
\] 
On the other hand, the entropy of the SBM $(C_2, M_2)$ which splits the big community into $q$ small ones of size $\frac{s}{q}$ and merges the $q$ small communities into one big is:
\[
S_2(c) = \mathrm{ln}\left[{s^2 + \sum_{i=1}^q m_i - 1 \choose \sum_{i=1}^q m_i}\right] + q^2\mathrm{ln}\left[{\frac{s^2 + c \times m_0}{q^2} - 1 \choose \frac{c \times m_0}{q^2}}\right]
\]

So, with $C_1 = \sum_{i=1}^q \mathrm{ln}\left[{\frac{s^2}{q^2} + m_i - 1 \choose m_i}\right]$ and $C_2 = \mathrm{ln}\left[{s^2 + \sum_{i=1}^q m_i - 1 \choose \sum_{i=1}^q m_i}\right]$, which are constants with respect to $c$:
\begin{align}
  S_1(c) - S_2(c) &= \mathrm{ln}\left[{s^2 + c \times m_0 - 1 \choose c \times m_0}\right] - q^2\mathrm{ln}\left[{\frac{s^2 + c \times m_0}{q^2} - 1 \choose \frac{c \times m_0}{q^2}}\right] + C_1 - C_2 \nonumber \\
  &= \mathrm{ln}\left[\prod_{k=1}^{c \times m_0} \frac{k + s^2 - 1}{k}\right] - \mathrm{ln}\left[\left(\prod_{k=1}^{\frac{c \times m_0}{q^2}} \frac{k + \frac{s^2}{q^2} - 1}{k}\right)^{q^2}\right] + C_1 - C_2 \nonumber \\
  &= \mathrm{ln}\left[\prod_{k=1}^{\frac{c \times m_0}{q^2}} \frac{\prod_{i=0}^{q^2 - 1}(k + s^2 - 1 + i \times \frac{c \times m_0}{q^2})}{(k + \frac{s^2}{q^2} - 1)^{q^2}}\right] + C_1 - C_2 \nonumber \\
  &> \mathrm{ln}\left[\prod_{k=1}^{\frac{c \times m_0}{q^2}} \left(\frac{k + s^2 - 1}{k + \frac{s^2}{q^2} - 1}\right)^{q^2}\right] + C_1 - C_2 \nonumber \\
  &> q^2 \sum_{k=1}^{\frac{c \times m_0}{q^2}} \mathrm{ln}\left[1 + \frac{(q^2 - 1)s^2}{q^2k + s^2 - q^2}\right] + C_1 - C_2
  \label{inequality}
\end{align}
Now, as
\[
\mathrm{ln}\left[1 + \frac{(q^2 - 1)s^2}{q^2k + s^2 - q^2}\right] \underset{k \rightarrow \infty}{\sim} \frac{(q^2 - 1)s^2}{q^2k + s^2 - q^2}
\]
and
\[
\sum_{k=1}^{\frac{c \times m_0}{q^2}} \frac{(q^2 - 1)s^2}{q^2k + s^2 - q^2} \underset{c \rightarrow \infty}{\rightarrow} \infty
\]
we have that
\begin{equation}
  q^2 \sum_{k=1}^{\frac{c \times m_0}{q^2}} \mathrm{ln}\left[1 + \frac{(q^2 - 1)s^2}{q^2k + s^2 - q^2}\right] \underset{c \rightarrow \infty}{\rightarrow} \infty
  \label{infinity}
\end{equation}
and thus, by injecting equation \ref{infinity} inside \ref{inequality}, $\exists c, \forall c' > c, S_2(c') < S_1(c')$. 
Which means that for any such pair of stochastic block models, there exists some density threshold for the big community in $C_1$ above which $(C_2, M_2)$ will be identified as the most likely model for all graphs $G \in \Omega_{(C_1, M_1)} \cap \Omega_{(C_2, M_2)}$.

\section{Consequences on model selection}
\label{sec_model_selection}

\begin{figure}[]
  \includegraphics[width=0.35\textwidth]{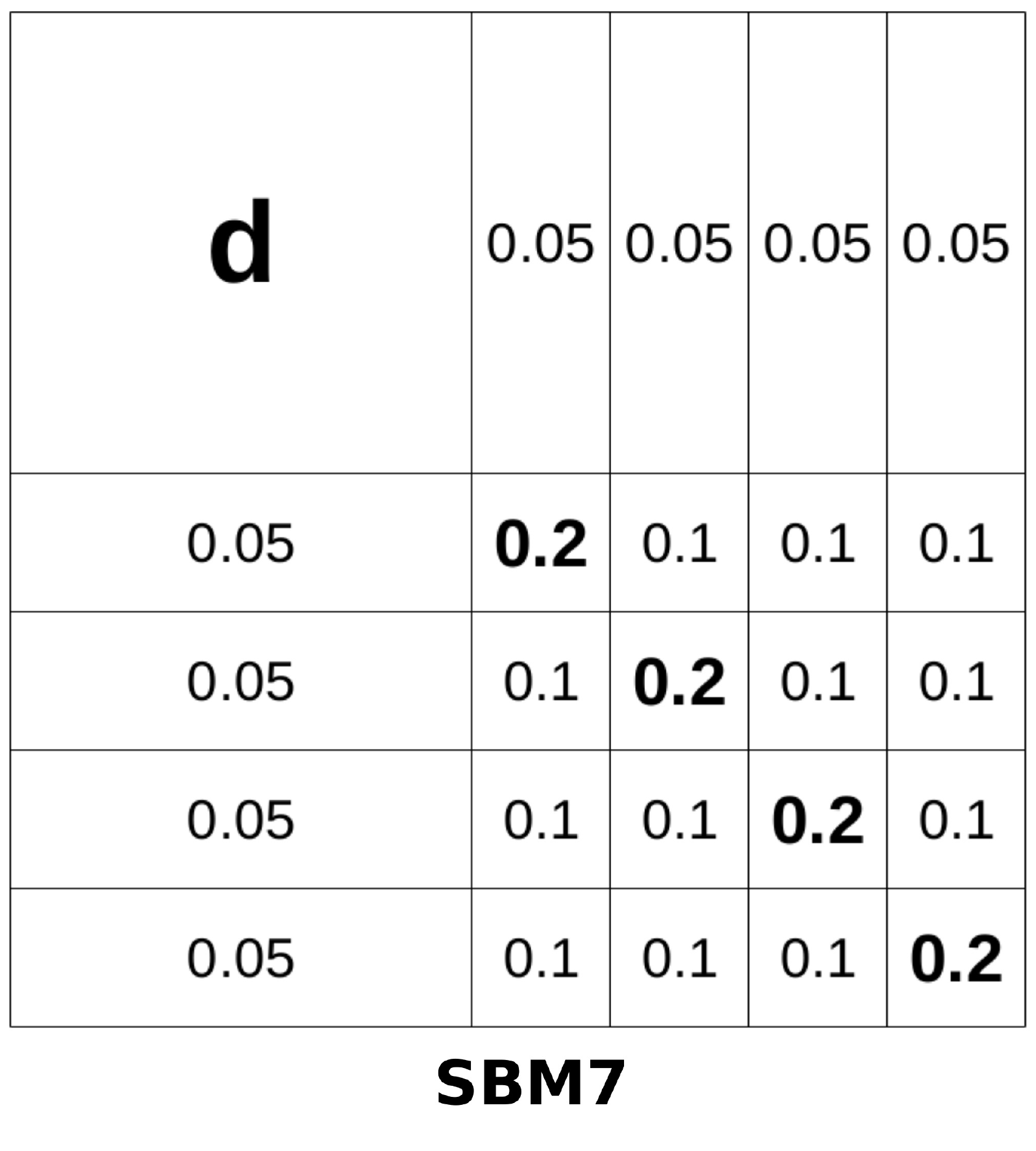}
  \hfill
  \includegraphics[width=0.6\textwidth]{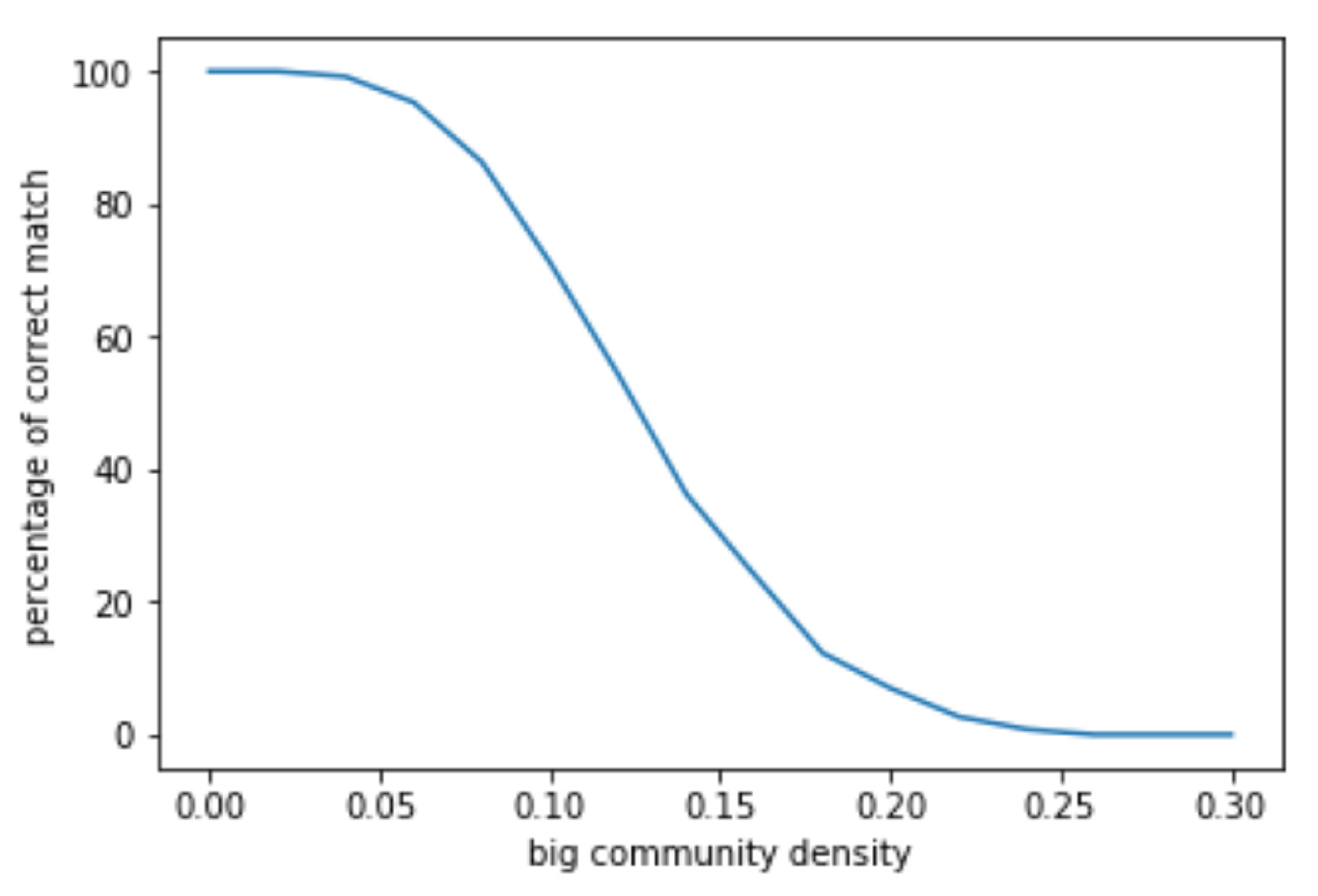}
  \caption{\label{SBM7} Block-to-block adjacency matrix of $SBM_7(d)$ (left) and percentage of graphs generated using $SBM_7(d)$ for which the original partition has a lower entropy than the inverted one against the density $d$ of the big community (right).}
\end{figure}

\begin{figure}[]
  \includegraphics[width=\textwidth]{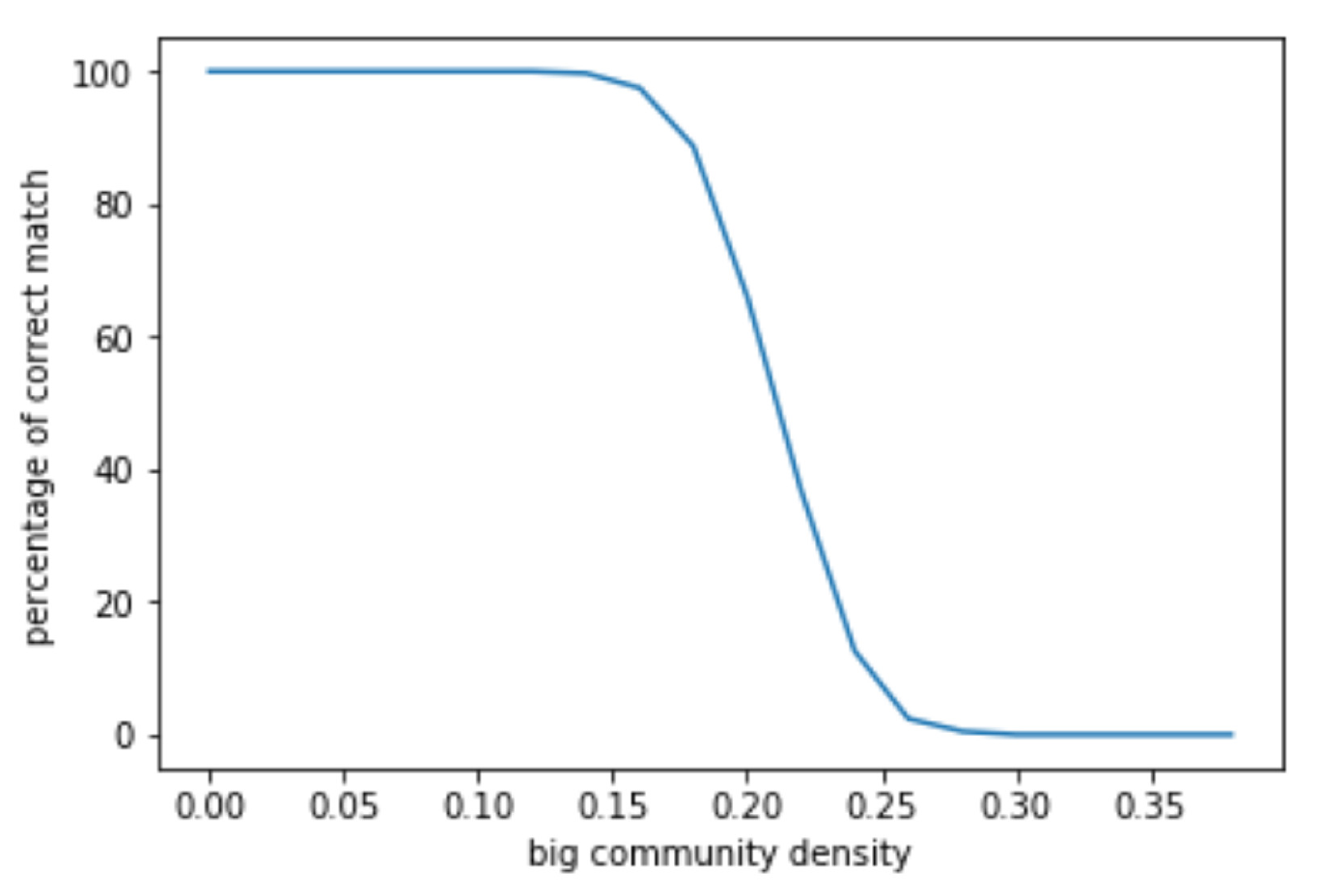}
  \caption{\label{SBM8} Percentage of graphs generated using $SBM_8(d)$ for which the original partition has a lower entropy than the inverted one against the density $d$ of the big community.}
\end{figure}

In practice, this phenomena implies that a model selection technique based on the minimization of entropy will not be able to identify correctly some SBM when they are used as generative models for synthetic graphs. To illustrate this, we generate graphs and try to recover the original partition. The experiment is conducted on two series of stochastic block models, one with relatively large communities and another one with smaller but more sharply defined communities:
\begin{itemize}
    \item $SBM_7(d)$ is made of 5 blocks (1 of 40 nodes, and 4 of 10 nodes). Its density matrix $D$ is given on figure~\ref{SBM7} (left) (one can deduce the block adjacency matrix by $M_{(c_i, c_j)} = |c_i||c_j| \times D_{(c_i,c_j)}$).
    \item $SBM_8(d)$ is made of 11 blocks (1 of 100 nodes, and 10 of 10 nodes). The internal density of the big community is $d$, it is 0.15 for the small ones and 0.01 between communities.
\end{itemize}
For each of those two models, and for various internal densities $d$ of the largest community, we generate 1000 random graphs. For each of these graphs, we compute the entropy of the original partition (correct partition) and the entropy of the partition obtained by inverting the big community with the small ones (incorrect partition). Then, we compute the percentage of graphs for which the correct partition has a lower entropy than the incorrect one and plot it against the density $d$. Results are shown on figure \ref{SBM7} and \ref{SBM8}.

We observe that as soon as $d$ reaches a given density threshold (about 0.08 for $SBM_7(d)$ and 0.18 for $SBM_8(d)$), the percentage of correct match drops quickly to 0. As $d$ rises over 0.25, the correct partition is never the one selected. It should be highlighted that in these experiments we only compared two partitions among the $B_n$ possible, so the percentage of correct match is actually an upper bound on the percentage of graphs for which the correct partition is identified. This means that if $SBM_7(d)$ or $SBM_8(d)$ are used as generative models for random graphs, with $d > 0.25$, and one wants to use bayesian inference for determining the original partition, it will almost never return the correct one. What is more, the results of section \ref{sec_density_threshold} show that this will occur for any SBM of the form described in figure \ref{SBM3}, as soon as the big community contains enough edges.

\section{Discussion}

We have seen in section \ref{sec_entropy_selection} that model selection techniques that rely on the maximization of the likelihood function to find the best node partition given an observed graph boils down to the minimization of the entropy of the corresponding ensemble of generable graphs in the microcanonical framework. Even in the case of bayesian inference, when a non-uniform prior distribution is defined on the set of possible partitions, entropy remains the criterion of choice between equiprobable partitions. Yet, as shown in section \ref{sec_high_density_zoom} and \ref{sec_density_threshold}, entropy behaves counter intuitively when a large part of the edges are concentrated inside one big community. In this situation, a partition that splits this community in small ones will have a lower entropy, even though the edge density is homogeneous. Furthermore, this happens even when the number and sizes of communities are known. Practically, as explained in section \ref{sec_model_selection}, this phenomena implies that stochastic block models of this form cannot be recovered using model selection techniques based on the mere minimization of the cardinal of the associated microcanonical ensemble.

Let's stress that contrary to the resolution limit described in \cite{fortunato2007resolution} or \cite{peixoto2013parsimonious}, the problem is not about being able or not to detect small communities with no prior knowledge about the graph, it occurs even though the number and sizes of communities are known. It is also different from the phase transition issue that has been investigated in \cite{decelle2011inference} \cite{decelle2011asymptotic} \cite{hu2012phase} \cite{abbe2015community} for communities detection or recovery because it happens even when communities are dense and perfectly separated. Entropy minimization fails at classifying correctly the nodes between communities because it only aims at identifying the SBM that can generate the lowest number of different graphs. A model which enforces more constraints on edge positions will necessarily perform better from this point of view, but this is a form of overfitting, in the sense that the additional constraints on edge placement are not justified by an heterogeneity in the observed edge distribution.

The results presented in this paper were obtained for a particular class of stochastic block models. First of all, they were obtained for the multigraph flavour of stochastic block models. As the node classification issue occurs also for densities below 1, they can probably be extended to simple graphs, but this would need to be checked, as well as the case of degree-corrected stochastic block models. Furthermore, the reason why the log-likelihood of a stochastic block model $C,M$ for a graph $G$ is equal to the entropy of $\Omega_{C,M}$ is that we consider the microcanonical ensemble, in which all graphs have an equal probability to be generated. It would be interesting to check if similar results can be obtained when computing $\mathbb{P}[G|C,M]$ in the canonical ensemble \cite{peixoto2012entropy}. Finally, we assumed that for a graph $G$ and two partitions $C_1$ and $C_2$ with the same number and sizes of blocks, the associated block-to-block adjacency matrices $M_1$ and $M_2$ have the same probability to be generated, and this assumption too could be questioned.

Yet, within this specific class of SBM, our results illustrate a fundamental issue with the stochastic block model statistical inference process. Since the random variable whose distribution we are trying to infer is the whole graph itself, we are performing statistical inference on a single observation. This is why frequentist inference is impossible, but bayesian inference also has strong limitations in this context. In particular, the only tool to counterbalance the observation and avoid overfitting is to specify the kind of communities we are looking for through the prior distribution. If it is agnostic about the distribution of edge densities among these communities, the mere minimization of the entropy of the posterior distribution fails to identify the heterogeneity in the edge distribution. Beside refining even more the prior distribution, another approach could be to consider a graph as the aggregated result of a series of edge positioning. If the considered random variable is the position of an edge, a single graph observation contains information about many of its realizations, which reduces the risk of overfitting.

\section*{Acknowledgments}
This work was supported by the ACADEMICS grant of the IDEXLYON, project of the Université de Lyon, PIA operated by \textbf{ANR-16-IDEX-0005}, and of the project \textbf{ANR-18-CE23-0004} (BITUNAM) of the French National Research Agency (ANR).

\bibliographystyle{unsrt}
\bibliography{minimum_entropy_stochastic_block_models}

\end{document}